\documentclass[final]{svjour2}
\usepackage{graphicx}
\usepackage{epsfig}
\makeatletter
\journalname{Journal of Low Temperature Physics}

\newcommand{\HeThree}{$^3$He}
\newcommand{\HeFour}{$^4$He}

\begin{document}

\newcommand{\hdblarrow}{H\makebox[0.9ex][l]{$\downdownarrows$}-}

\title{A glassy contribution to the heat capacity of hcp $^4$He solids}

\author{Jung-Jung Su$^1$ \and Matthias J. Graf$^1$ \and Alexander V. Balatsky$^{1,2}$}

\institute{1: Theoretical Division, Los Alamos National Laboratory, Los Alamos, New Mexico 87545, USA
\\2: Center for Integrated Nanotechnology, Los Alamos National Laboratory, Los Alamos, New Mexico 87545, USA}

\date{\today}

\maketitle
	
\begin{abstract}

We model the low-temperature specific heat of solid $^4$He in the hexagonal closed packed structure by invoking 
two-level tunneling states in addition to the usual phonon contribution of a Debye crystal 
for temperatures far below the Debye temperature, $T < \Theta_D/50$. 
By introducing a cutoff energy in the two-level tunneling density of states, 
we can describe the excess specific heat observed
in solid hcp $^4$He, as well as the low-temperature linear term in the specific heat. 
Agreement is found with recent measurements of the temperature behavior of both specific heat and pressure. 
These results suggest the presence of a very small fraction, at the parts-per-million (ppm) level, of two-level tunneling systems in solid $^4$He,
irrespective of the existence of supersolidity.  
\end{abstract}

\keywords{solid $^4$He \and glass \and supersolid \and quantum phase transition}
\PACS{67.80.B-, 64.70.Q-, 67.80.bd}

\section{Introduction}
The anomalous frequency and dissipation behavior seen in torsion oscillators (TO) \cite{Chan04} at low temperatures has stimulated numerous investigations, since it was suggested to be the signature of supersolidity \cite{Andreev69,Chester67,Reatto69,Leggett70,Anderson84}. Recent successive TO experiments \cite{Rittner06,Kondo07,Aoki07,Clark07,Penzev07,Hunt09} 
confirmed the finding of the anomalous behavior.
In addition,  hysteresis behavior and long equilibration times have been observed
\cite{Aoki07,Hunt09,Kim09}, 
which depend strongly on growth history and annealing \cite{Rittner06}. In the same temperature range, transport experiments including shear modulus \cite{Beamish05,Beamish06}, ultrasonic \cite{Goodkind02,Burns93} and heat propagation \cite{Goodkind02} have shown various anomalous behaviors. However, no clear sign in mass flow 
\cite{Greywall77,Paalanen81,Beamish05,Beamish06,Sasaki06,Ray08,Bonfait89,Balibar08} 
and structural measurements \cite{Burns08,Blackburn07} emerged that can demonstrate the occurrence of a phase transition.

It has been anticipated that thermodynamic measurements will resolve the existing controversy, since any true phase transition should be accompanied by a thermodynamic signature. The search for such signatures proved to be challenging by 
the experiments conducted so far,  including measurements of the specific heat
\cite{Swenson62,Frank64,ClarkChan05,LinChan07,LinChan09,WestChan09}, 
pressure dependence of the melting curve \cite{Todoshchenko06,Todoshchenko07}, 
and pressure-temperature measurements of the solid \cite{Grigorev07,Grigorev07b}. 
The main difficulties lie in measuring small signals at low temperatures in the presence of large backgrounds. 
With improving experiments at low temperatures, measurements down to 20 mK were conducted.
While there is still no clear evidence of transition in the melting curve experiments, recent pressure measurements and specific heat measurements have both shown deviations from the expected pure Debye lattice behavior. 
Balatsky and coworkers \cite{Balatsky07,Graf08} 
argued that the deviations occurring at low temperatures might be related to a glass phase, where the role of two-level systems could be taken by tunneling dislocation loop segments. This description is also accompanied by an argument that the excess entropy associated with the deviation is too small for supersolidity to lead to detectable mechanical effects, if it is due to a
supersolid fraction alone.  

In this paper, we expand the above glass model to describe the glass freezing transition occurring in the thermodynamic experiments of ultrapure $^4$He solid. We model the subsystem of tunneling dislocation segments with a compact 
distribution of two-level excitation spacings, see Fig.~\ref{fig:DOS}. 
We can then describe the behavior of a transition as well as the zero-temperature extrapolation of the glassy 
behavior seen in thermodynamic measurements. 
Our results show that the low-temperature deviations in the measured specific heat can be explained by contributions from a 
glassy fraction of the solid. 
These results add further support to our previous interpretation of torsion oscillator experiments in terms of a backaction 
due to a glassy subsystem, which is possibly present in solid \HeFour\ \cite{Graf09,Graf08,Nussinov07}.

\section{Glass model for the specific heat}

\begin{figure} 
\begin{center}
\includegraphics[width=.6\linewidth,angle=0,keepaspectratio]{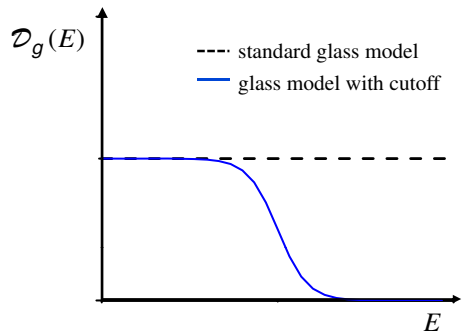}
\end{center}
\caption{Density of states (DOS) of the two-level tunneling system. The black-dash line represents the DOS for the standard glass model \cite{Anderson72,Phillips72,Balatsky07}, 
while the blue-solid line is the truncated DOS used in this work describing a two-level system with a cutoff energy. 
}\label{fig:DOS}
\end{figure}

We propose a thermodynamic model to describe the measured low-temperature specific heat.
We postulate the existence of a distribution of two-level tunneling (TLS) systems in solid hcp $^4$He. 
These TLS may be created through complex configurations of dislocation loops embedded in the crystal or strain during growth.
The distributions of length and number of dislocation segments depend on $^3$He concentration, quenching and other growth 
processes. 
In this paper, we are comparing the effect of different growth processes on ultrapure \HeFour\ containing at most (nominally)
1 ppb of \HeThree\ impurities. At such low levels of impurities, we expect to see the intrinsic properties of solid
\HeFour.

We start with the expansion of the standard TLS model (Fig.~\ref{fig:DOS}). 
  In the standard glass model \cite{Anderson72,Phillips72,Balatsky07} for solids, 
the density of the TLS states is assumed to be constant,  ${\cal D}(E)$ = const., to account for the linear temperature 
coefficient in the specific heat at low temperatures. 
However, it is  well-known in the context of conventional glasses that a more careful analysis of specific heat data
gives rise to a power law deviating slightly from linearity.
The deviation  may be attributed to phonon relaxation processes \cite{Zimmermann81}
or to a density of states (DOS) of the TLS that is not constant over a characteristic
energy $E_c$ of level spacings \cite{Lasjaunias75,Lasjaunias78}. 
Here, we neglect the time dependence in the specific heat due to phonons scattering off from the TLS
and for simplicity concentrate on the cutoff dependence of the DOS at high energies. 
For example, such a cutoff could be due 
to the finite barrier height of double-well potentials giving rise to the TLS,
because in real materials the tunneling barrier has 
an upper bound set by lattice and dislocation configurations \cite{Jaeckle72}. 
This is also the reason why glassy behavior is usually 
only seen at sufficiently low temperatures. At high temperatures the thermal energy can easily overcome the barrier 
and the TLS effectively become a system of noninteracting single oscillators. 
At $T < \Theta_D/50$, the specific heat of solid \HeFour\ is well described by
\begin{eqnarray}
C(T) &=& C_{L}(T) + C_{g}(T) ,
\end{eqnarray}
where the phonon contribution to the molar specific heat is given by $C_L(T) = B_L T^3$,  with coefficient
$B_L = 12 \pi^4 R/5 \Theta_D^3$, $R=8314$ mJ/(mol K) is the gas constant, 
and $\Theta_D$ is the Debye temperature. 
The second term describes the glass contribution due to the TLS subsystem  and is given by
\begin{eqnarray} \label{C_glass}
C_{g} (T) = k_B R \frac{d}{d T} \int_0^\infty dE \, E \, {\cal D}_{g}(E) \, f(E)  ,
\end{eqnarray}
with $k_B$ being the Boltzmann constant and $f(E)$ is the Fermi function. 
The DOS of the TLS may be modeled by the box distribution function
\begin{eqnarray} \label{DOS}
{\cal D}_{g} (E) = \frac{1}{2}{\cal D}_0 \left[ 1-\tanh((E-E_c)/W) \right]  .
\end{eqnarray}
Here ${\cal D}_0$ is the zero-energy DOS per energy, $E_c$ is a characteristic cutoff energy, 
and $W$ is the width of the truncated density of states.
For $E_c \to \infty$, one obtains the standard hallmark result of glasses at low temperatures:
\begin{eqnarray}
C_{g}(T) = B_g T,
\end{eqnarray} 
where $B_g=k_B R {\cal D}_0$. As we will elaborate in the next section, the glass coefficient $B_g$ has an intrinsic finite value at low temperature even for the purest $^4$He samples,
independent of \HeThree\ concentration.

\section{Results and discussion}

\begin{figure} 
\begin{center}
\includegraphics[width=0.9\linewidth,angle=0,keepaspectratio]{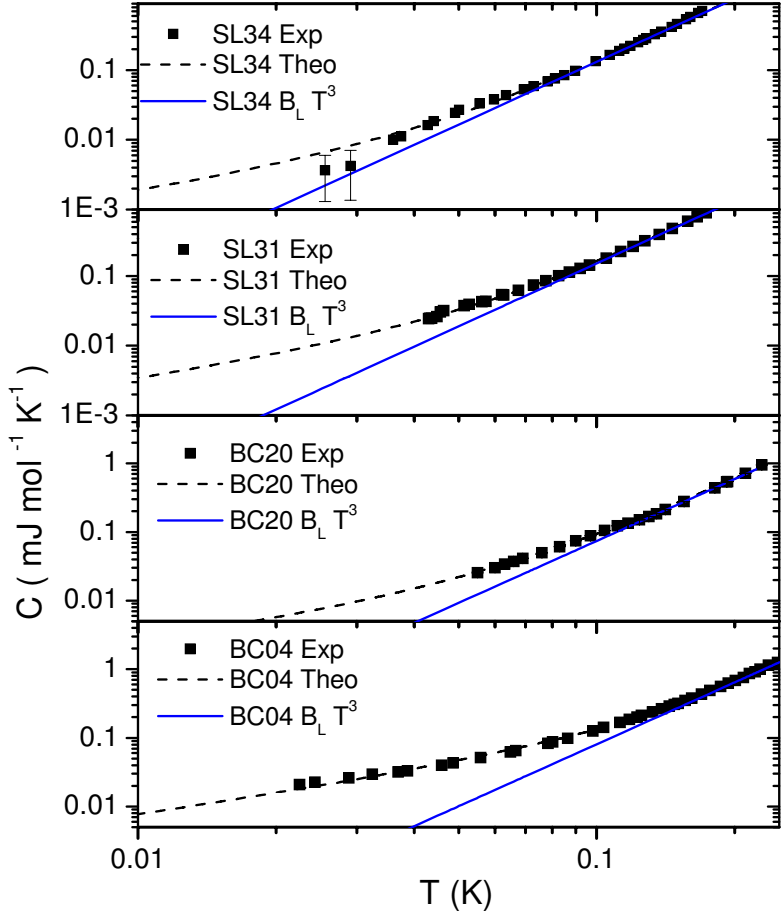}
\end{center}
\caption{The specific heat $C$ of solid \HeFour\ with nominally
1 ppb \HeThree\ for four different growth processes
taken from Lin et al. \cite{LinChan09}. 
The experimental data (squares) are fit by $B_L T^3$ at high temperatures $T>0.16$ K (blue line). 
The black-dash line shows the total calculation with both lattice and glass contributions. 
}\label{fig:Clog}
\end{figure}

\begin{figure}
\begin{center}
\includegraphics[width=0.9\linewidth,angle=0,keepaspectratio]{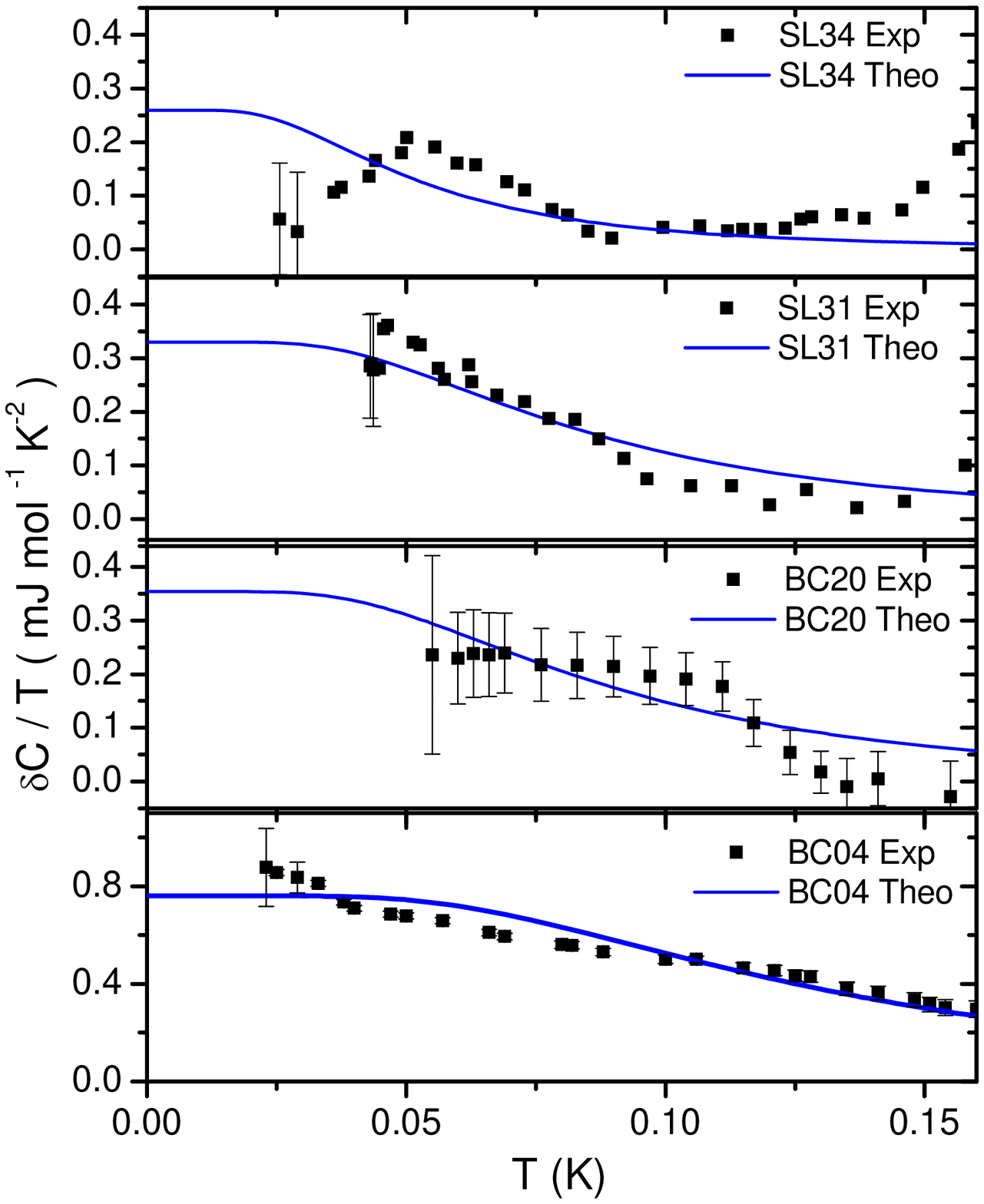}
\end{center}
\caption{$\delta C/T$ for experiments (squares) and the modified glass model with a cutoff energy 
in the TLS DOS (blue line) for same four samples shown in Fig.~\ref{fig:Clog}, 
with $\delta C = C - C_L$. 
The error bars are obtained from Ref.~\cite{LinChan09}.
}\label{fig:CovT}
\end{figure}
\subsection{Specific Heat}
  We compare our model calculations with experimental data by the Penn State group \cite{LinChan09,LinChan07} for four different 
growth processes in ultrapure \HeFour\ with at most 1 ppb of \HeThree\ impurities. 
We re-analyzed their experiments and obtained quite different assignments for the state of two of their samples:
We find two solid-liquid coexistence samples with solid ratios 0.34 (SL34) and 0.31 (SL31) \cite{countmole},
while we agree with the assignments for the samples grown with the blocked capillary method (BC).  One was grown
over 20 hours (BC20), the other one was grown over 4 hours (BC04). 
Notice that sample SL34 corresponds to the 75\% solid-liquid coexistence sample 
reported by the Penn State group and that the SL31 sample corresponds to their constant pressure sample (CP).
The experimental data are described with three parameters: ${\cal D}_0$, $E_c$ and the Debye temperature $\Theta_D$. 
We first determine $\Theta_D$, or the lattice contribution, from the high-temperature data (see Fig.~\ref{fig:Clog}). 
The phonon contribution is then subtracted from $C$ to obtain the difference $\delta C = C - C_L$. 
We fit $\delta C/T$ with our specific heat formula in Eq.(~\ref{C_glass}) for a glass.

We show the difference in specific heat over temperature, $\delta C/T$, for four different growth processes in 
Fig.~\ref{fig:CovT}. 
The model describes well the low-temperature part for all four cases and the high-temperature part is within the scatter
of the experimental data. In these plots we fixed the width of the cutoff to $W = 1 \, {\rm \mu eV}$. With $W \ll E_c$, 
we have verified that there is no qualitative difference when varying $W$ within reasonable ranges.
Notice that the shape of $\delta C(T)$ depends strongly on the subtraction of the high-temperature phonon contribution. 
Although the Debye temperature used for the BC samples is reasonable, it is not unique in determining the shape of 
$\delta C(T)$, since a usual 1\% uncertainty in the data of $C(T)$ at high temperatures leads to quite large uncertainties
in $\delta C(T)/T$.

\begin{table}[h!]
\begin{center}
\begin{tabular}{c|ccccccc}
& $P$ & $V_m$ & $\Theta_D $ & ${\cal D}_0 \times 10^4$ & $E_c \times 10^2$ 
&$n_{\rm TLS}$ & $\Delta S$\\ 
& (bar) &(cm$^3$/ mol)& (K) & (1/meV) & (meV) & (ppm) & ($\mu$J/(mol K)) \\
\hline 
SL34      & 25 & 21.25 & 24.5 & 2.2 & 1.7 & 3.7 & 21.3\\
SL31      & 25 & 21.25 & 24.8 & 2.9 & 2.2 & 6.4 & 36.9\\
BC20      & 33 & 20.46 & 29.7 & 3.0 & 2.3 & 6.9 & 39.5\\
BC04      & 33 & 20.46 & 28.9 & 6.5 & 3.3 & 21.5& 115.0\\
\end{tabular}
\end{center}
\caption{Physical and model parameters: Debye temperature $\Theta_D$, zero energy TLS DOS
${\cal D}_0$, cutoff energy $E_c$, concentration of TLS $n_{\rm TLS}$ and excess entropy $\Delta S$. }
\end{table} \label{para_table}

The physical and model parameters of the four samples grown under different conditions are summarized in 
Table~1.
The Debye temperature $\Theta_D$ of $^4$He increases linearly with decreasing molar volume $V_m$ in the range 
between 21 and 16 cm$^3$/mol \cite{Edwards65,Greywall71}. 
This is also the pressure range in which the experimental data were taken.
We find that $\Theta_D$ for both BC grown samples agrees reasonably well with literature values of
28-29 K at molar volume $V_m=20.46$ cm$^3$/mol and pressure $P=33$ bar. The same consistency is found for the coexistence samples, 
after correcting for the solid-liquid ratio \cite{chanerratum}, 
with the reported literature value of $\Theta_D \sim 25$ K at $P=25$ bar \cite{Edwards65,Greywall71}. 
A caveat is warranted regarding sample SL31. Since it has a slightly larger concentration of TLS than SL34, although it should be equally pure. Hence, one needs to consider that it was supposedly grown under CP conditions at $P=38$ bar. 
However, it is not clear if the final pressure of the cell was actually 25 bar or higher, nor how strained the crystal was when the initial pressure of 38 bar was lost. All these experimental unknowns may change our assignment for the solid-liquid coexistence ratio for SL31. Thus its solid-to-liquid ratio of $x=0.31$ should be considered a lower bound \cite{chanerratum}.

The glassy behavior is mainly characterized by the zero-energy DOS and cutoff energy of the TLS, which are both noticeably larger in BC04 than in the others. This may be explained by a rapid growth process creating a strained crystal, which gives rise to both a larger TLS concentration and a smaller cutoff energy, i.e., a smaller maximum tunneling barrier height. 
On the other hand, the comparison of sample SL31 with BC20 leads us to believe that $^3$He concentration does not appear to play an important role at the 1 ppb concentration level and below.
The coexistence sample SL31 is supposed to be purer than BC20 ($^3$He atoms are more soluble in liquid than solid \HeFour). Our analysis does not show significant differences between parameters for both cases. This is supported by our analysis that the TLS concentration of these samples ranges from $3.7$ to $21.5$ ppm, which is at least 1000 times larger than the nominal $^3$He concentration. We argue that this small concentration of \HeThree\ impurities, compared to the intrinsic concentration of TLS in even the best crystals grown by the Penn State group, explains the experimental finding that the sample grown with the BC method for over 4 hours and 0.3 ppm of \HeThree\ impurities exhibits nearly identical behavior as sample BC04 with only 1 ppb of \HeThree. 
We therefore conclude that the intrinsic property of solid $^4$He was measured in the samples with 1 ppb  \HeThree\ impurities, 
given the much larger TLS concentration levels found.

\subsection{Entropy Analysis}
We also calculated the excess entropy,
\begin{equation}
\Delta S(T) = \int_0^T dT' \,\delta C(T')/T' ,
\end{equation}
associated with the excess specific heat due to the low-temperature transition or crossover into a glass phase. 
The advantage of an entropy analysis over that of the specific heat lies in the robustness and simplicity
of counting states in an equilibrium phase compared to detailed model calculations for the specific heat.
We find consistently for specific heat experiments \cite{ClarkChan05,LinChan07,LinChan09} that the
obtained values are 5 to 6 orders of magnitude smaller compared to the theoretical prediction for a supersolid
if the entire sample underwent Bose-Einstein condensation (BEC).  In the limit of a non-interacting BEC one finds
$\Delta S_{BEC}=15/4 (\zeta(5/2)/\zeta(3/2))\, R (T/T_c)^{3/2}\sim (5/4)\, R\sim 10.4$ J/(K mol). 
This means that if $\Delta S$ is indeed due to supersolidity, then the supersolid volume fraction is at most 
11 ppm or 0.0011\% in the most disordered or quenched sample of the four ultrapure samples studied in this work,
i.e., sample BC04.  
Such a supersolid fraction in the specific heat
of bulk \HeFour\ is more than 100 to 1000 times smaller than is usually reported for the 
non-classical rotational inertia fraction (NCRIF) in torsion oscillator experiments. 
This enormous discrepancy between supersolid fractions in specific heat and torsion oscillator experiments was already 
noticed in Refs.~\cite{Balatsky07,Graf08}.
Until to date, this discrepancy remains a major puzzle that is hard to reconcile within a purely supersolid scenario. 

\begin{figure}
\bigskip
\vskip -0.3 truecm
\begin{center}
\includegraphics[width=1.2\linewidth,angle=0,keepaspectratio]{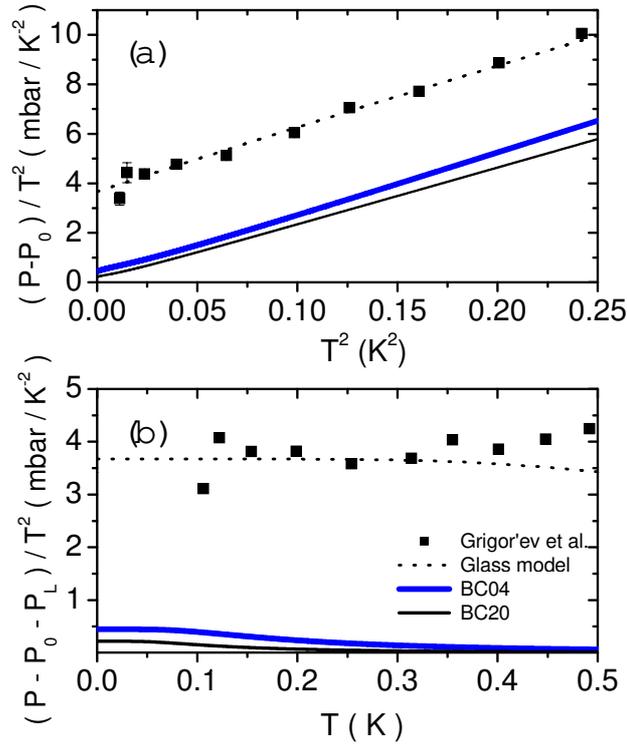}
\end{center}
\vskip -0.3 truecm
\caption{
(a) ($P-P_0)/T^2$ vs. $T^2$ for $P \sim 33$ bar. The squares represent the data reported by Grigor'ev et al. \cite{Grigorev07,Grigorev07b} The black dotted line is the model calculation with $D_0=4.95 \times 10^{-3}$(1/meV), $E_c=0.2$meV, $\Theta_D=28.6$K and the same $W$ as that in the specific heat calculation. The blue-thick and black-thin lines are predictions for BC04 and BC20, respectively. The predicted curves for BC04 and BC20 use the same parameters as those shown in Fig.~\ref{fig:Clog}. 
(b) ($P-P_0-P_L)/T^2$ vs. $T$ for the same data collection as in (a). This plot clearly shows that the Grigor'ev et al. data is characterized by a much larger $D_0$ and $E_c$(roughly seven times larger) than those in the Lin et al. data.    
}\label{fig:Pressure}
\end{figure}

\subsection{Comparison with Pressure Measurement}
Next we compare the specific heat with the pressure measurements and their deviations from a perfect Debye lattice 
behavior. 
We are using thermodynamic relations between the specific heat and pressure.
The quantities to characterize the pressure measurement in the combined lattice and glass models 
are $a_L$ and $a_g$ defined by
\begin{eqnarray}
P(T) \equiv  P_0+P_L(T)+P_g(T) = P_0 + a_L T^4+ a_g T^2  ,
\end{eqnarray}
where $P(T)$ is the pressure at temperature $T$. 
$P_0$, $P_L$, $P_g$ are the corresponding pressure contributions of the ions at zero temperature, lattice vibrations,
and two-level  excitations of the glass. 
The Mie-Gr\"{u}neisen theory gives the thermodynamic relation between pressure and specific heat 
\begin{eqnarray}
\left(\frac{\partial P}{\partial T} \right)_V = 
\frac{\gamma_g}{V_m} \, C_{g,V} + \frac{\gamma_L}{V_m} \, C_{L,V} ,
\end{eqnarray}
where $\gamma_i$ are the Gr\"uneisen coefficients of the glass excitations ($g$) and of the lattice vibrations ($L$).
Literature values for the Gr\"uneisen coefficient of phonons in solid hcp $^4$He range between $2.6 < \gamma_L < 3.0$ 
\cite{Grigorev07b,Driessen86}, while nothing is known about $\gamma_g$ of glassy $^4$He. 
For simplicity we assume in our calculation
of $P(T)$ based on our model of the specific heat that $\gamma_g \sim \gamma_L =2.6$.

In Fig.~\ref{fig:Pressure}(a) we show the temperature dependence of the pressure data $(P-P_0/T^2)$ reported by Grigor'ev et al. \cite{Grigorev07,Grigorev07b} as well as those predicted for samples BC04 and BC20 from specific heat measurements using a glass description. The glass contribution utilized in modeling the pressure data is given by
 \begin{eqnarray}
 \frac{P_g (T)}{T^2} = \frac{\gamma_g}{V_m} k_B R \int_0^{\infty} dE\, E
{\cal D}_g (E) f(E) ,
 \end{eqnarray}
with the TLS density of states $D(E)$ defined in Eq.~(\ref{DOS}).
All curves show finite intercepts with the ordinate, which we attribute
to the glassy contribution on top of the $T^2$-term due to lattice
vibrations.
The glassy contribution can be shown most clearly by plotting
$(P-P_0-P_L)/T^2$ vs.\ $T$ (see Fig.~\ref{fig:Pressure} (b)). The glass contribution is larger in Grigor'ev's sample and survives to higher temperatures compared to predictions for samples BC04 and BC20.
The parameters $D_0$ and $E_c$ are roughly seven times bigger in Grigor'ev's pressure data than predicted from the specific heat data by Lin et al.
This may be expected for a much more rapidly grown solid. It is worth
mentioning that the Gr\"uneisen coefficient $\gamma_g$ affects the result
in the same way as $D_0$. The coefficient $\gamma_g$ can actually vary significantly from one sample to another\cite{Ackerman1984}. Here, for simplicity, it is taken to be the same both in pressure and in specific heat measurements.
To summarize, we have shown that the pressure measurement as well as the specific heat measurement are consistently described in the framework of the modified glass model. The higher TLS concentration extracted from Grigor'ev's data compared to Lin's data is likely due to a much faster cooling rate or due to a
 larger Gr\"ueneisen coefficient or both. 

\section{Conclusions}

We have shown that a simple TLS model explains quantitatively the linear temperature coefficient in the specific heat and
the low-temperature deviation of the specific heat, $\delta C$, from that of a perfect Debye crystal without postulating 
the existence of supersolidity.  By truncating the TLS DOS above a characteristic cutoff energy $E_c$, 
the bump-like feature in $\delta C$ can fairly well be captured within the scatter and uncertainty of the data. 
Furthermore, our thermodynamic analysis results in Debye temperatures  $\Theta_D$ and TLS DOS ${\cal D}_0$ 
that are in qualitative agreement with known $P(T)$ measurements in the solid \cite{Grigorev07,Grigorev07b}. 
This suggests the possibility 
of a glassy subsystem at the ppm level in hcp $^4$He crystals.
The presence of a glassy subsystem is consistent with recent reports of long relaxation times in 
torsion oscillators \cite{Hunt09}, $P$ vs. $T$ measurements \cite{Grigorev07,Grigorev07b}, and transport measurements
 \cite{Goodkind02,Burns93}. 
In order to uniquely determine the ground state of solid \HeFour, i.e., whether it exhibits a supersolid 
or glass transition, more accurate measurements of the specific heat and thermal conductivity
at lower temperatures and up to 0.5 K are needed. 
Finally, relaxation time measurements of heat pulses could provide useful and much needed information about 
the dynamics of a possible glass or supersolid subsystem in hcp \HeFour\ solid.

\begin{acknowledgements}
We enjoyed valuable discussions with Z. Nussinov, J.C. Davis, B. Hunt, E. Pratt, J.M. Goodkind, 
and J. Beamish. We thank M.H.W. Chan and J. West for discussion and for sharing their data.
This work was supported by the US Dept.\ of Energy at Los Alamos National Laboratory
under contract No.~DE-AC52-06NA25396.
\end{acknowledgements}

%
%

%

\begin{thebibliography}{10}
\bibitem{countmole} We calculate the molar specific heat from the specific heat of sample SL34 by obtaining the number of 
mole of solid \HeFour\ in the solid-liquid coexistence sample of Ref.~\cite{LinChan09}. 
First, we extract from the data the slopes of $C/T^3$ of the solid-liquid phase and the pure liquid at 
high temperatures, $B_{SL}=2.31$ mJ/K and $B_{\rm Liq}=0.72$ mJ/K.
Next, we get from literature the slope for the solid at the melting line, where solid and liquid coexist, 
$B_{\rm Sol}=5.45$ mJ/K at $P=25$ bar and $V_m=21.25$ cm$^3$/mol \cite{Greywall78,Edwards65}. 
Finally, we calculate the molar ratio of the solid from the two-phase mixing relation:
$B_{SL}= x B_{\rm Sol}+(1-x) B_{\rm Liq}$ with $x=0.34$ and the mole number of the solid 
$(0.93$ cm$^3$)/(21.25 cm$^3$mol$^{-1}$)=0.0149 mol.
\end{thebibliography}
\end{document}